# SURVEYING UNDRERGRADUATE GREEK STUDENT'S KNOWLEDGE ON SOME ISSUES OF RADIATIONS AND NUCLEAR ENERGY

[1]Mirofora Pilakouta[1] and John Sinatkas[2]

[1]University of West Attica, Department of Industrial Design & Production
P. Ralli and Thivon 250, 12244 Egaleo, Greece

[2]University of Western Macedonia at Kastoria, Department of Informatics,
P.O. Box 30, GR-521 00, Kastoria, Greece

## Abstract

In this work we present some results from a survey aimed to assess the knowledge and views of the Greek undergraduate students (technology oriented) on some issues of radiations, nuclear energy and their consequences.

Findings indicate that the, examined group, of Greek students have a series of misconceptions and faulty views on radiations and general nuclear issues. No significant differences in the students responses related to the type of secondary school they attended were found. Moreover, analysis according to gender, indicated that females are less informed than males in most of the examined issues

## Introduction

The beneficial applications of nuclear physics and ionizing radiation (medical applications, nuclear energy, industrial applications) have become nowadays a part of our everyday life. Consequently it is important for general public (GP) to have a basic understanding on these issues. This will improve their perceptions about the benefits and potential risks associated with these applications.

This knowledge is normally imparted in secondary schools. Therefore, for the non - major in physics/chemistry undergraduate students and for most of the general population, the formal information about these issues ends after finishing the secondary school.

In literature, there are studies from different countries that explore the views, conceptions and general knowledge about issue related to "radiation" of mainly high school students . Older studies examined student conceptions mainly in the area of nuclear -ionizing radiation (Eijkelhof et al. 1990) (Millar, K. Klaassen, and Eijkelhof 1990), (Millar and Gill 1996). More recent studies (Cooper, Yeo,

---

Previous affiliation

[1]Technological Education Institute of Piraeus 250, Thivon Ave., Aigaleo GR-122 44 Athens-Greece

[2]Technological Educational Institute of Western Macedonia at Kastoria, P.O. Box 30, GR-521 00, Kastoria, Greece

and Zadnik 2003), (Rego and Peralta 2006), (Plotz 2017), (Neumann 2014) (Kontomaris and Malamou 2017) examine issues in several radiation bands. The main findings on misconceptions are summarized and discussed in (Neumann and Hopf 2012) (Plotz 2017) . The general conclusion is that there is poor knowledge and confusion in many issues related to radiation ionize and non-ionized. Students as well as public still have a number of lay-ideas on these issues which are different from scientific ideas (Eijkelhof et al. 1990).

The confusion is partly attributed, to the non - scientifically valid way the media transmit these issues as well as to social environment. B.A. Sesen and E.Ince (Sesen and Ince 2010) registered several scientifically incorrect conceptions about ionizing radiation in internet sources that students used for gathering information.

Furthermore, it is underlined the low contribution of the educational system, and the need for suitable modernized curricula, to motivate students to be more aware on radiations and nuclear issues (Rego and Peralta 2006) (Cooper et al. 2003)(Eijkelhof et al. 1990) (Millar, Klaassen, and Eijkelhof 1990).

From our experience, our students - in Technological Educational Institutes in Greece – have also limited and confused knowledge. These subjects are discussed in a limited extent in the Greek secondary school and the knowledge and opinion forming on these issues for both, public and students, is highly effected from non-formal sources i.e. the media and the Internet as well.

In our country hasn't been yet any extended study covering these issues. A Greek study (Χαρτζάβαλος Σ. 2009) on scientific literacy in basic issues of nuclear physics in first-year physics students, revealed that only 20% of students responded satisfactorily. A preliminary survey aimed at the assessment of the knowledge and perceptions of our undergraduate first year engineering students on radiations (case study), radioactivity and nuclear applications have been conducted by the author (Pilakouta 2011) after Fukushima accident. The findings guided the development of new activities (Pilakouta, Savidou, and Vasileiadou 2017) as well as other educational material like Educational videos (Πηλακούτα and Βαρσάμης 2013) and seminars. Furthermore another survey was conducted 2012-2017 with an extended questioner including questions on non-ionizing radiations as well. In the new survey the source of student's knowledge in these issues was also examined.

Some results of this survey are presented here. These results help us to create and enrich our informational material, in popularized form, for Greek students and General Public. Only for curiosity, we tried also to see whether our data on student's performance reflected the knowledge of a public group consisted mainly from the administrative staff of TEI Piraeus.

**Methodology**

The research questions' guiding this study was:
- What do the Greek, technology oriented undergraduate students, know about these issues?
- Does gender or prior education of the participants affect their knowledge and views

*Survey Design - Sample*

Having in mind the effectiveness and the results of a preliminary investigation (a small questionnaire - 8 questions - in the form of diagnostic tests, that explored students' knowledge of TEI Piraeus (Pilakouta 2011), the questionnaire was reformed and enriched with additional questions of general interest related to the issues of nuclear and general radiation applications.

Most of the questions are based on what our experience registered as common misconceptions, some of them were proposed by some experts that test the questionnaire and also some ideas from relative studies were used (Rego and Peralta 2006)(Millar et al. 1990) (Cooper et al. 2003).

The questionnaire consisted of three parts. The first part included questions that answered by Yes, No, Do not know in order to get a general sense of the participants knowledge. The second part included multiple choice questions with specific answers in order to have more analytical data. Finally the third part included demographic questions, data for the source of the participant information as well as students' attitude for seminars related to issues of general interest.

The questionnaire was administered to the students and the public group mainly via mail. A number of students answered the questionnaire in the physics lab.

The questions are listed in tables 1 and 2. The issues examined through questions A1, A2, A3, B1 and B2, are considered as issues included in the Greek General Lyceum curriculum. The issues in the rest of the questions (A4, A5, B3, B4, B5) are considered as issues of general interest.

The collected data were analyzed using SPSS. Their frequencies were mainly recorded and the possible statistical differences between selected subgroups were investigated.

The students sample was a convenience sample consisted of 313 undergraduate students, 254 from TEI Piraeus and 59 from TEI of West Macedonia (Kastoria). Overall the students were mainly on the first year of their studies.

- Previous education of the participants: 77% students came from General lyceums and 23% from "Vocational Lyceums". In Greece there are two types of upper secondary schools the "General Lyceums" (GL) which are academically oriented, and the "Vocational Lyceums" (EPAL) which are vocationally oriented.
- Gender of the participants: in the sample there are 34% female and 66% male.

Another convenience sample consisted of 131 Greek citizen's (public employs' mainly from the administrative staff of TEI Piraeus). In this sample 53 % were male and 47% females. The majority 82% were university graduates and 18% secondary school graduates. The age of the public participants was between 25-65 years old.

**Results and discussion**

Below we present results from the analysis of the student's responses totally, by gender, and by the type of Lyceum they had attended. More over some results from public are also presented for comparison.

Overall, almost in all the questions, figure: 1, and 2, Table1 and 2, the percentage of the correct responses in all groups and subgroups were low (less than 50%) or very low (less than 20%). The only exception was the answers in the questions A1 and A2.

**Table:1 Questions of part A**

| | |
|---|---|
| A1 | Have you ever heard of natural radioactivity? |
| A2 | Do all types of radiation cause the same effects in the human body? |
| A3 | Do Laser devises emit X-rays |
| A4 | Someone who takes a thyroid scintigraphy medical exam is considered carrying radioactive elements? |
| A5 | Nuclear power plants in standard operation, emit a great less amount of pollutants than carbon power plants? |

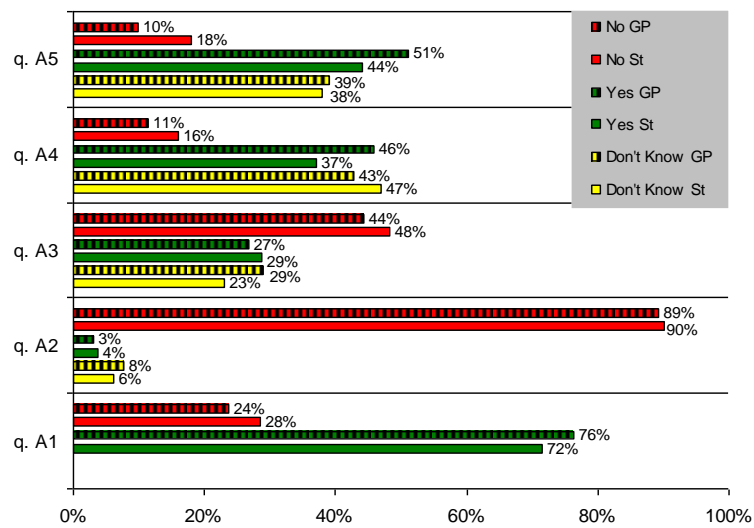

Figure 1: Comparison between total responses of the students and the public in the selected questions of A part

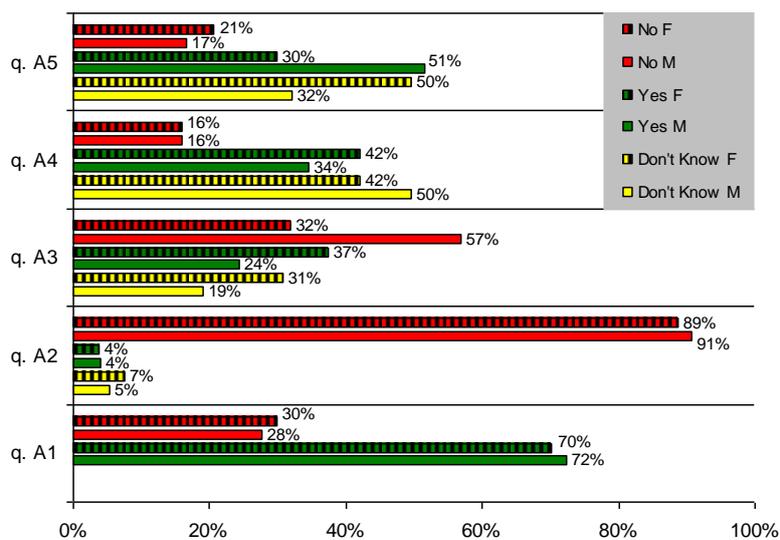

Figure 2: Results % from student's responses with respect to gender

*Results from Students responses*

The answers to questions A1 and A2, figure: 1, shows that about 72% of the students respondents were aware for the existence of natural radioactivity and a majority of 90% of them know that not "all the types of radiation cause the same effects in the human body".

In question A3 about one out of 4 students (23%) don't know whether Laser devices emits X-rays and about 29% think that they do which is indicative of the confusion of the «origin of X radiation»

Overall, only 37 % of the students in question A4 know that during the thyroid scintigraphy the patient carries radioactive elements. Also, a lot of them 47% admit that they don't know. This shows how inadequate is the knowledge about this (and not only) application of nuclear radiation in medicine.

In Question A5 only 44% of the students are informed that Nuclear power plants in standard operation emit less amount of pollutants than carbon power plants and a rather high percentage 38% doesn't know.

In part B, questions B1 and B2 examines the knowledge in some types of radiations. These questions have the same alternative answers (γ rays, X rays, cell phone radiation, radio waves, visible light and don't know) and accepted more than one answer. From fiqure:3, we may notice that the majority, of the students were most aware that X-Rays and gamma rays could cause in a high dose genetic mutation but from data analysis follows that only 29% of the participant students, table 2, indicate the correct answer that is exclusively x and γ rays. Moreover the fact that 30% of student thinks that cell phone radiation and radio waves 18% or even the visible light may cause genetic mutation, demonstrates a rather limited knowledge on the differences between the various types of radiation (ionizing and non ionizing radiation) and indicate confusion about the radiations that may cause genetic mutation. Furthermore, these results are in contrast with the 90% of them (fiqure 1, Q.A2) that they respond that not all the types of radiation cause the same effects in the human body.
Responses in question B2 indicate that students do not know clearly which of the suggested radiations are emitted by radioactive nucleus. Overall 64% of them indicated gamma rays together with other radiations, 36% indicated X Rays and other radiations, but only 28% of the participants indicated the correct answer that is exclusively γ rays.

This finding in conjunction with the choices they made in the available answers, ie: some of them (10%) think that cell phones radiation and radio waves (16%) are emitted by unstable nucleus, leads to the conclusion that their knowledge about radiations is not only poor but is also very confused.

**Table 2 Questions of part B**

**Q. B1** Which of the following radiation/s could cause in a high dose genetic mutation* :

| γ- rays and x-rays [a] | **Students** | Male | Female | **General population** |
|---|---|---|---|---|
| | 29% | 33% | 22% | 21% |

**Q. B2** Which of the following radiations are emitted by radioactive nucleus*

| γ-rays [a] | **Students** | Male | Female | **General population** |
|---|---|---|---|---|
| | 28% | 31% | 22% | 22% |

**Q. B3** Which of the factors listed below contribute more to the total dose of ionizing radiation received by the average person during his life?

| | Natural [a] radiation | Nuclear accidents | Nuclear medicine | Nuclear industry | Don't know |
|---|---|---|---|---|---|
| **Students** | 18% | 20% | 35% | 5% | 22% |
| Male | 20% | 18% | 36% | 4% | 22% |
| Female | 14% | 22% | 34% | 7% | 23% |
| **General population** | 28% | 22% | 22% | 4% | 24% |

**Q. B4** Two of the main radioactive elements which are likely to be released into the atmosphere due to an accident in a nuclear power plant are:

| | Iodine [a] Cesium | Iodine Uranium | Uranium Cesium | Uranium Plutonium | Don't know |
|---|---|---|---|---|---|
| **Students** | 10% | 9% | 8% | 50% | 24% |
| Male | 11% | 9% | 10% | 48% | 23% |
| Female | 8% | 9% | 5% | 54% | 24% |
| **General population** | 22% | 8% | 10% | 41% | 20% |

**Q. B5** About how many people died because of high radiation dose in the first 3 months after the nuclear accident at Chernobyl in 1986?

| | 5-50 [a] | 50-100 | > 10.000 | > 100.000 | Don't know |
|---|---|---|---|---|---|
| **Students** | 12% | 10% | 21% | 19% | 39% |
| Male | 14% | 9% | 20% | 17% | 40% |
| Female | 8% | 10% | 24% | 22% | 36% |
| **General population** | 5% | 14% | 29% | 11% | 41% |

[a] Correct answer, *The percentage in each of the available answers in questions B1 and B2 for the students, are presented in fiqure:2

Examining the results from question B3: "Which of the following factors contribute more in the total amount of ionized radiation absorbed by the average human during his life" Table 2, we notice that, only (18%) of the students knows that the main contribution to human irradiation comes from natural radioactivity. Overall 35%, believe that the main source of irradiation comes from medical examinations, 25% from nuclear plants or nuclear accidents and 22% admit that they do not know. Comparing this result with the responses in Q.A1 in figure 1, where 72% of the student were aware for the existence of natural radioactivity, we conclude that they almost know only the term of natural radioactivity without knowing in what extent affect our life.

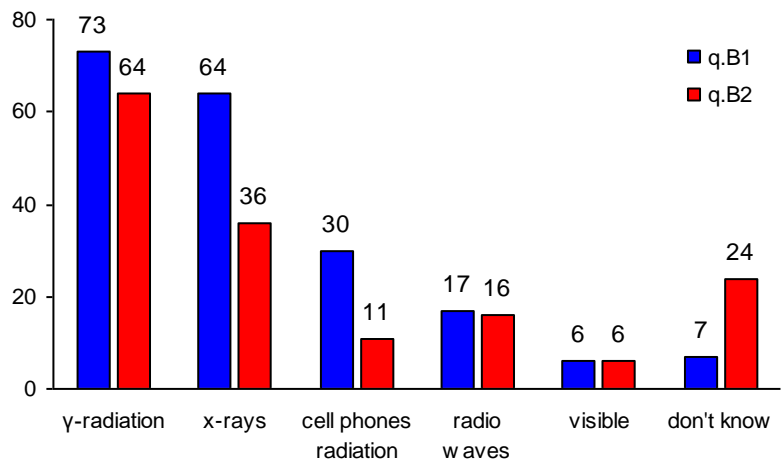

Figure 3: Answers on questions B1 and B2. Each column gives the percentage of answers identifying that type o radiation in each question

The belief that the contribution of medical examinations is higher than the contribution of other sources, may related to the widespread use of these kind of examinations in our country.

Question B4 and B5 are related to the nuclear accidents.

B4: Which are the two basic radioactive elements that are possible to be emitted in the atmosphere due to an accident at a nuclear power plant?

Only 10% answered correctly (cesium and iodine). Half of students (50%), table 2, have the wrong perception that Plutonium and Uranium are the radioactive elements that are mainly released in the environment after a nuclear accident. Some of them think of iodine and Uranium 10%, another 8% thinks of cesium and Uranium and 24% declared that they don't know. This indicates that the majority of the student confuses the nuclear fuel with the reaction products.

B5: How many individuals lost their life due to high radiation at the first 3 months after the Chernobyl incident at 1986

This question was selected in order to confirm the influence of the media to the development of wrong impressions.

Only 12% of the students knew that less than 50 deaths can be attributed directly to the radioactive contamination due to the Chernobyl accident. The majority of them either declare that they don't know (39%), or believe that deaths were more than 10.000 (21%), more than 100.000 (19%) or about 50-100 (9%). The above result shows the huge misconception, most of the students (and general population) have about the direct and the aftermath effects of radioactive contamination. The main reason for this faulty impression is that the media and several organizations (UN, Atomic Energy Agency, WHO) report mainly the potential effects of the ionizing radiation (wikipedia)(Cooper et al. 2003).

*Gender and prior education differences in students responses*

Among the participants in the students group, male students had a slightly higher score of correct answers than their female counterparts. The percentage of correct answers, fig.2, fiq.3 and table 2, have statistically significant differences (significance 5%) in questions A3, A5, and B2, B1. Thus mainly in the "school" related questions.

Examining the data with respect the former education of the students, GL and EPAL, we notice only a slight but not significant higher performance for the students of GL in almost all questions. Overall only in two questions the difference was statistically significant. In question B2 the students from GL gave better results but in question B4 the students from EPAL had a better performance. Having in mind that such issues rarely are discussed in EPAL, these findings indicates again, the poor contribution of our school to the students knowledge about these serious issues.

*Students versus selected group considered as general public*

Although the sample that is considered as general public is not representative (convenience sample), we just comment some differences with our students responses

In figure: 1 the results from students and GP % responses in the question of part one are presented. Having in mind that the majority 82% of the convenience public was university graduates, it is obvious that misinformation about radiation and nuclear issues crosses all educational levels.

The results show that, in the questions related to the secondary school curriculum there is no significant difference between the performances of the two groups (significance level 5%). A one-tailed test shows that GP has higher percentage of correct answers in the Questions A4, B3, and B4. In question B5 (although the very low number of correct answers >12%) students have higher score. Thus we may say that GP has a better knowledge on these general interest issues than the students.

*Which is the primary source of student's information about radioactive issues?*

Overall, one out of five students 20 % indicated school as the primary source of their information about the above issues figure: 4a. The majority 52 % has information from internet, 15% from mass media (radio-TV) and a small number of students (13%) have other sources of information. This distribution underlines the poor contribution of secondary school to the knowledge about these serious issues. Moreover shows that about 67% of the students use informal sources for their information.

*Students' attitude for seminars related to issues of general interest*

The majority of the students, 65% (see Figure: 4b), state (definitely and rather yes) that they would participate in seminars on issues of general interest (like the seminar on radiations and nuclear issues), 27% said maybe and only 8% said probably no or no,

This indicates a positive attitude towards the general interest seminars, ie after secondary school they prefer a non- traditional, but valid way to be informed in these issues. This is interesting and we should enhance it because it is an important aspect of social learning since it links the educational system with the wider community.

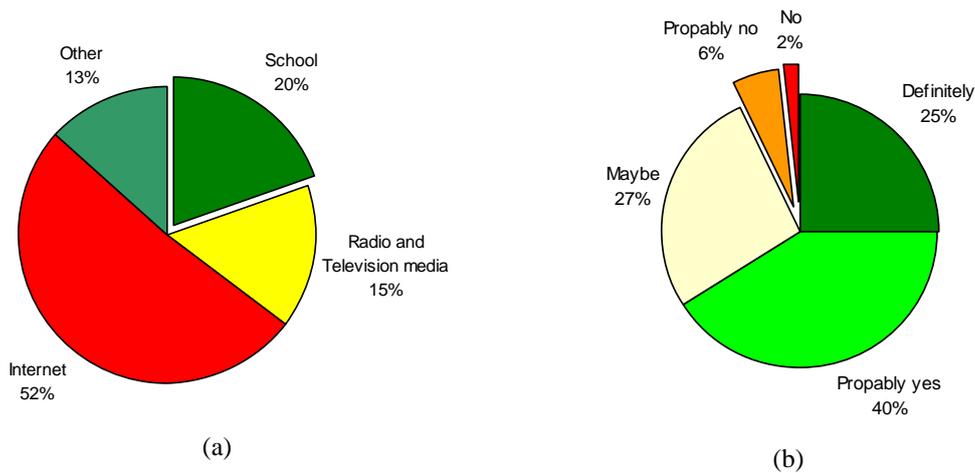

Figure 4 a) The primary source of students information about radioactive issues
b) Students' attitude for the seminars related to issues of general interest

It should be noted that the GP potential interest for general interest seminars was roughly analog to that of the students.

## Summary - Conclusions

Some results from a survey aimed to assess the knowledge and views of a convenience group of Greek undergraduate students (technology oriented) on some issues of radiations, nuclear energy and their consequences were presented.

The results, confirmed what experience registered ie, that the examined group of Greek students have false beliefs and perceptions on issues of nuclear and radiation issues and their applications, although these issues touch our everyday life and have great social impact. In short, most students have heard something about radiation and know something about the origin of the several radiations, but not in a level to differentiate them and assess the type of dynamic danger from them. Furthermore, although there is particular concern on nuclear accidents, only few of them know something about the function of the nuclear plants, the type of main pollutants of high concern after an accident and the extent of the immediately effects . These findings are valuable and would guide our effort to develop and disseminate informational material for our students in an effort to increase the knowledge and reduce the confusion.

Although we believe that the gender differences in the knowledge in science is narrowed, in this survey we've notice that the percentage of females that answered correctly most of the questions was lower than that of males.

Finally, both groups (students and citizens) are interested in and wish to have information on relevant issues. The students show a highly positive attitude for general interest seminars and this is something that we must have in mind. We should find ways to improve students' knowledge and to motivate their interest, for the useful applications of radiations and nuclear physics in our life as well as for the harmful effects of radioactivity in health and environment.

To get secure assessment of the Greek student's and public knowledge about the referred issues, a more systematic and representative survey is needed.